\documentclass[sn-mathphys]{sn-jnl}

\usepackage{amsmath}
\usepackage{graphicx}
\usepackage{gensymb}
\usepackage{caption}
\usepackage{subcaption}
\usepackage{bm}

\DeclareMathOperator*{\argmin}{arg\,min}

\jyear{2021}%

\theoremstyle{thmstyleone}%
\theoremstyle{thmstyletwo}%

\theoremstyle{thmstylethree}%

\begin{document}

\title[\; ]{A marginal modelling approach for predicting wildfire extremes across the contiguous United States}


\author[1]{\fnm{Eleanor} \sur{D'Arcy}}\email{e.darcy@lancaster.ac.uk}
\equalcont{These authors contributed equally to this work.}

\author*[1]{\fnm{Callum J. R.} \sur{Murphy-Barltrop}}\email{c.barltrop@lancaster.ac.uk}
\equalcont{These authors contributed equally to this work.}

\author[2]{\fnm{Rob} \sur{Shooter}}\email{robert.shooter@metoffice.gov.uk}
\equalcont{These authors contributed equally to this work.}

\author[3]{\fnm{Emma S.} \sur{Simpson}}\email{emma.simpson@ucl.ac.uk}
\equalcont{These authors contributed equally to this work.}

\affil[1]{\orgdiv{STOR-i Centre for Doctoral Training}, \orgname{Lancaster University}, \orgaddress{\street{Fylde Avenue}, \city{Lancaster}, \postcode{LA1 4YR}, \country{United Kingdom}}}

\affil[2]{\orgdiv{Department}, \orgname{Met Office Hadley Centre}, \orgaddress{\street{FitzRoy Road}, \city{Exeter}, \postcode{EX1 3PB}, \country{United Kingdom}}}

\affil[3]{\orgdiv{Department of Statistical Science}, \orgname{University College London}, \orgaddress{\street{Gower Street}, \city{London}, \postcode{WC1E 6BT}, \country{United Kingdom}}}


\abstract{This paper details a methodology proposed for the EVA 2021 conference data challenge. The aim of this challenge was to predict the number and size of wildfires over the contiguous US between 1993 and 2015, with more importance placed on extreme events. In the data set provided, over 14\% of both wildfire count and burnt area observations are missing; the objective of the data challenge was to estimate a range of marginal probabilities from the distribution functions of these missing observations. To enable this prediction, we make the assumption that the marginal distribution of a missing observation can be informed using non-missing data from neighbouring locations. In our method, we select spatial neighbourhoods for each missing observation and fit marginal models to non-missing observations in these regions. For the wildfire counts, we assume the compiled data sets follow a zero-inflated negative binomial distribution, while for burnt area values, we model the bulk and tail of each compiled data set using non-parametric and parametric techniques, respectively. Cross validation is used to select tuning parameters, and the resulting predictions are shown to significantly outperform the benchmark method proposed in the challenge outline. We conclude with a discussion of our modelling framework, and evaluate ways in which it could be extended.}




\maketitle

\section{Introduction}\label{sec:intro}
\subsection{Motivation and data description}
This paper details an approach to the data challenge organised for the EVA 2021 conference. The subject of the challenge was wildfire modelling, and two important sub-challenges were proposed within this setting. In particular, teams were asked to develop methods for predicting the number of fires (i.e., individual fires that are separated in space), as well as the amount of burnt land resulting from these fires, over different months for gridded locations across the continental United States (US). 

In the absence of mitigation, wildfires can have devastating consequences, including loss of life and damage to property. The northern California wildfire in October 2017 burned approximately 150,000 acres of land, resulting in 7,000 damaged structures and 100,000 evacuations~\citep{wong2020}. Recent increases in both the number and severity of wildfires can be linked to climate change, and in particular to anthropogenic warming \citep{Jones2020}. Focusing specifically on the western US, \cite{Zhuang2021} demonstrate that a high proportion of the observed increases in weather events leading to wildfires may be attributed to this aspect of climate change. Extreme events in wildfire modelling are especially important; the more individual wildfires that occur, the greater the potential destruction, and the impact of large wildfires (in terms of the amount of land area burnt) can be particularly devastating. It is therefore of interest to develop models for wildfires, and in particular wildfire extremes.

The challenge data set consists of monthly wildfire count (CNT) and burnt area (BA) observations from 1993 to 2015 at 3,503 grid cell locations spanning the contiguous US. There are $35$ auxiliary variables also recorded relating to land cover types, climate and altitude. Observation locations are arranged on a $0.5\degree \times 0.5\degree$ (approximately 55 km $\times$ 55 km) regular grid of longitude and latitude coordinates, with observations recorded from March to September; further details are provided by \cite{Opitz2021}.

In order to compare the predictions produced by the teams participating in the data challenge, several observations were removed from the data to act as a validation set; this contained 80,000 observations for each of CNT and BA. The selection of these validation points was not done completely at random, so there is some spatio-temporal dependence between them. This will be discussed further in Section~\ref{subsec:validation}, with a pictorial example given in Figure~\ref{fig:validationLocations}. Let $CNT_i$ and $BA_i$, $i=1,\dots,N$, denote the $i$-th observation of the wildfire CNT or BA data, respectively, where $N=563,983$ is the total number of observations across the training and validation sets for each variable over all sites, months and years. We denote the set of observation indices in the validation sets for CNT and BA by $CNT^{val},BA^{val}\subset\{1,\dots,N\}$, respectively, with $\vert CNT^{val}\vert=\vert BA^{val}\vert=80,000$. An important feature is that the validation indices are not identical for the CNT and BA data, but there is a reasonable overlap, i.e., $CNT^{val}\neq BA^{val}$ but $CNT^{val}\cap BA^{val}\neq\emptyset$. We discuss ways to exploit this aspect in Section~\ref{subsec:dataFeatures}.

The objective of the challenge was to predict cumulative probability values for both CNT and BA at the times and locations in their respective validation sets. The resulting estimates were then ranked using a score computed from the true observed values, with lower scores corresponding to more accurate probability predictions. These scores were weighted so that more importance is placed on the estimation of the extremes; see \cite{Opitz2021}. Statistical techniques that do not explicitly model the tail are therefore unlikely to produce the best scores.

\subsection{Data exploration}\label{subsec:data_exploration}
In this section, we give an overview of features of the data set that motivate our modelling approach. We consider the relationship between CNT and BA, as well as the temporal non-stationarity of each variable separately; we also investigate how these features vary over the spatial domain.

We begin by exploring the dependence between CNT and BA; for the bulk of the data, we consider Kendall's $\tau$ measure of rank correlation, whilst for the extremes we consider the widely-used measures $\chi$ and $\bar\chi$. Consider a random vector ($X,Y$) with marginal distribution functions $F_X$ and $F_Y$, respectively. \cite{Coles1999} define $\chi=\lim_{u\rightarrow1}\chi(u)$, where $\chi(u)=\Pr(F_Y(Y)>u \mid F_X(X)>u)\in[0,1]$, as a measure of asymptotic dependence. If $\chi\in(0,1]$, $X$ and $Y$ are said to be asymptotically dependent, with $\chi=1$ corresponding to perfect dependence. Asymptotic independence between $X$ and $Y$ is present only when $\chi=0$, meaning that $\chi$ fails to signify the level of asymptotic independence. To account for this, \cite{Coles1999} define a further measure that provides additional detail in this case, namely $\bar\chi=\lim_{u\rightarrow1}\bar\chi(u)\in(-1,1]$ where \begin{equation*}
    \bar\chi(u)=\frac{2\log\Pr(F_Y(Y)>u)}{\log\mathbb{P}(F_Y(Y)>u, F_X(X)>u)}-1.
\end{equation*} Under asymptotic dependence, $\bar\chi=1$, and for asymptotic independence, $\bar\chi<1$; the further sub-cases $\bar\chi\in(0,1)$ and $\bar\chi\in(-1,0)$ correspond to positive and negative association, respectively, while $\bar\chi=0$ indicates independence.

We estimate these measures separately for subsections of the US to investigate spatial variability in the dependence structure between CNT and BA. We start by splitting the spatial domain into quadrants corresponding to the north east (NE; $>37.5^\circ$N, $<100^\circ$W), south east (SE; $\leq37.5^\circ$N, $<100^\circ$W), south west (SW; $\leq37.5^\circ$N, $\geq100^\circ$W) and north west (NW; $>37.5^\circ$N, $\geq100^\circ$W). Kendall's $\tau$ measure suggests strong overall correlation between CNT and BA, with estimates of $0.926~(0.925,0.927)$, $0.827~(0.825,0.829)$, $0.858~(0.855,0.860)$ and $0.868~(0.867,0.870)$ for the NE, SE, SW and NW respectively, with the values in brackets denoting 95\% confidence intervals obtained via bootstrapping. However, estimates of $\chi(u)$ and $\bar{\chi}(u)$ suggest this dependence diminishes in the extremes, leading to asymptotic independence. We obtain estimates (and 95\% bootstrap confidence intervals) of $\chi(0.999)=0.071 \hspace{.1em}(0.043,0.126),$ $0.038  \hspace{.1em}(0.017,0.072),$ $0.012  (0,0.024)$ and $0.043 (0.024,0.077)$, and $\bar\chi(0.999)=0.438 (0.343,0.521),$ $0.282 (0.191,0.392),$ $0.092 (-0.05,0.179)$ and $0.317 (0.253,0.413)$, for the NE, SE, SW and NW regions respectively. The NE region exhibits the strongest dependence between CNT and BA in the bulk of the data, as well as the strongest extremal dependence. We extended this analysis to look at smaller spatial domains, but our conclusions did not change. 

\begin{figure}[!ht]
\centering
\begin{subfigure}{.5\textwidth}
    \centering
    \includegraphics[width=\textwidth]{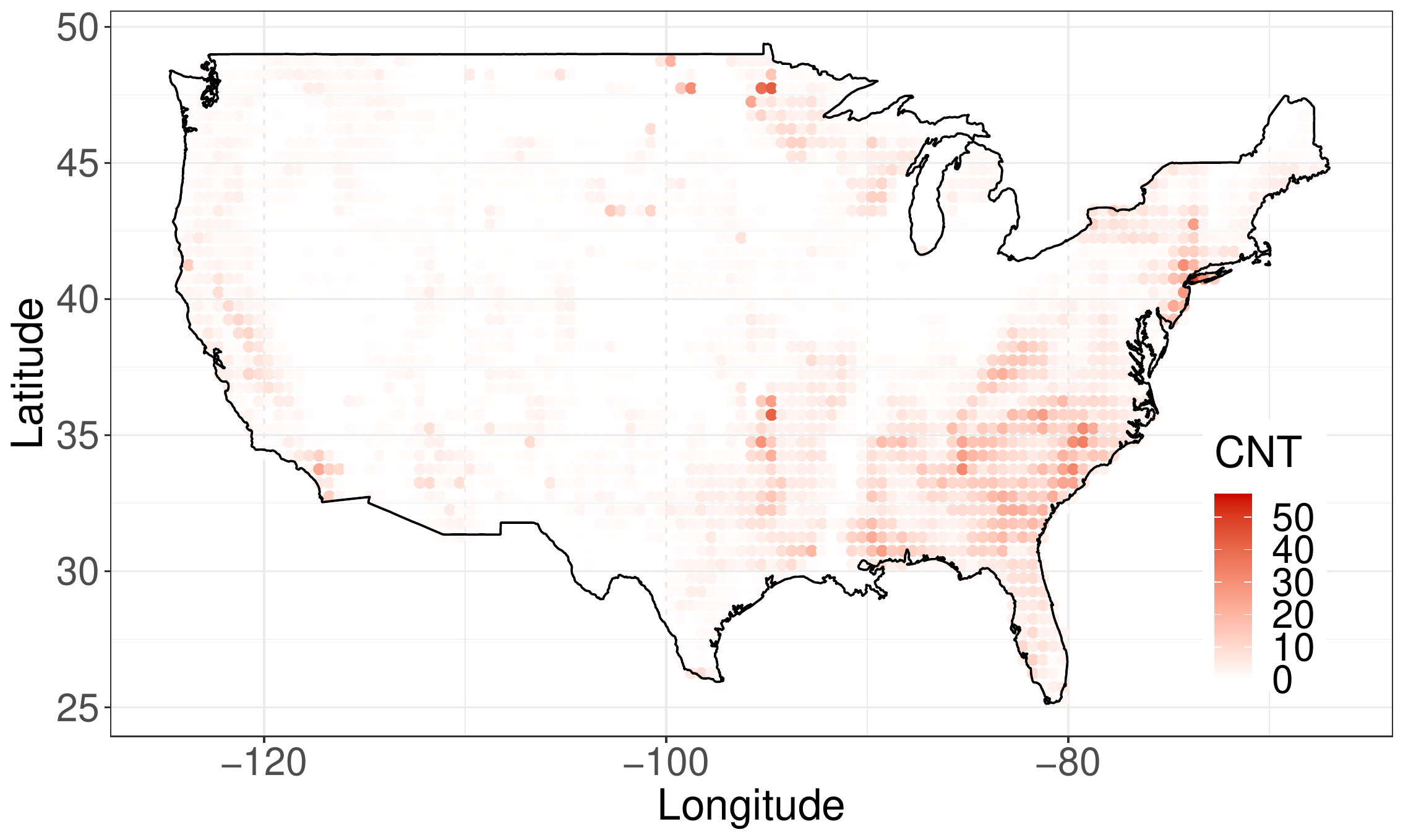}
    \caption{}
\end{subfigure}%
\begin{subfigure}{.5\textwidth}
    \centering
    \includegraphics[width=\textwidth]{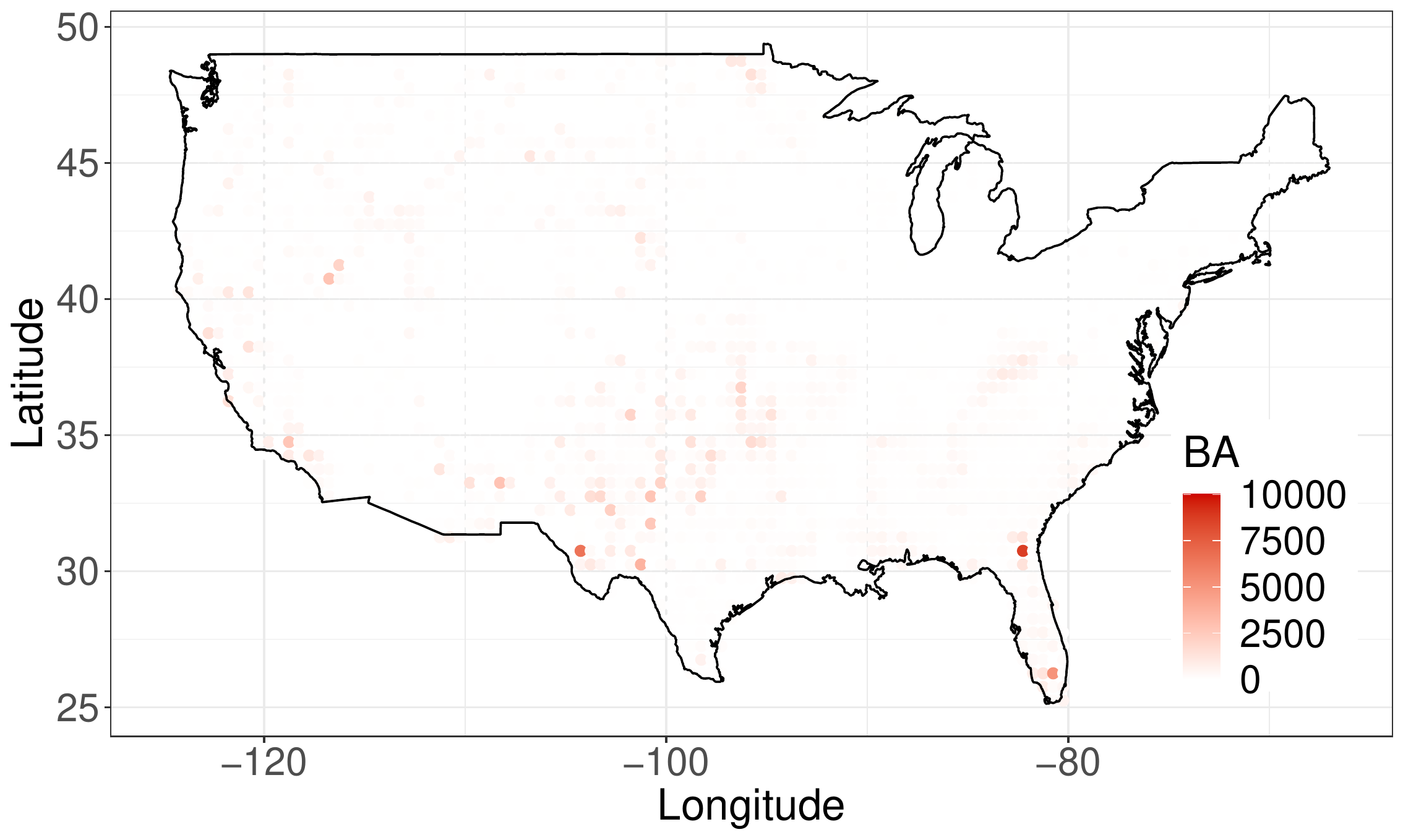}
    \caption{}
\end{subfigure} \\
\begin{subfigure}{.5\textwidth}
    \centering
    \includegraphics[width=\textwidth]{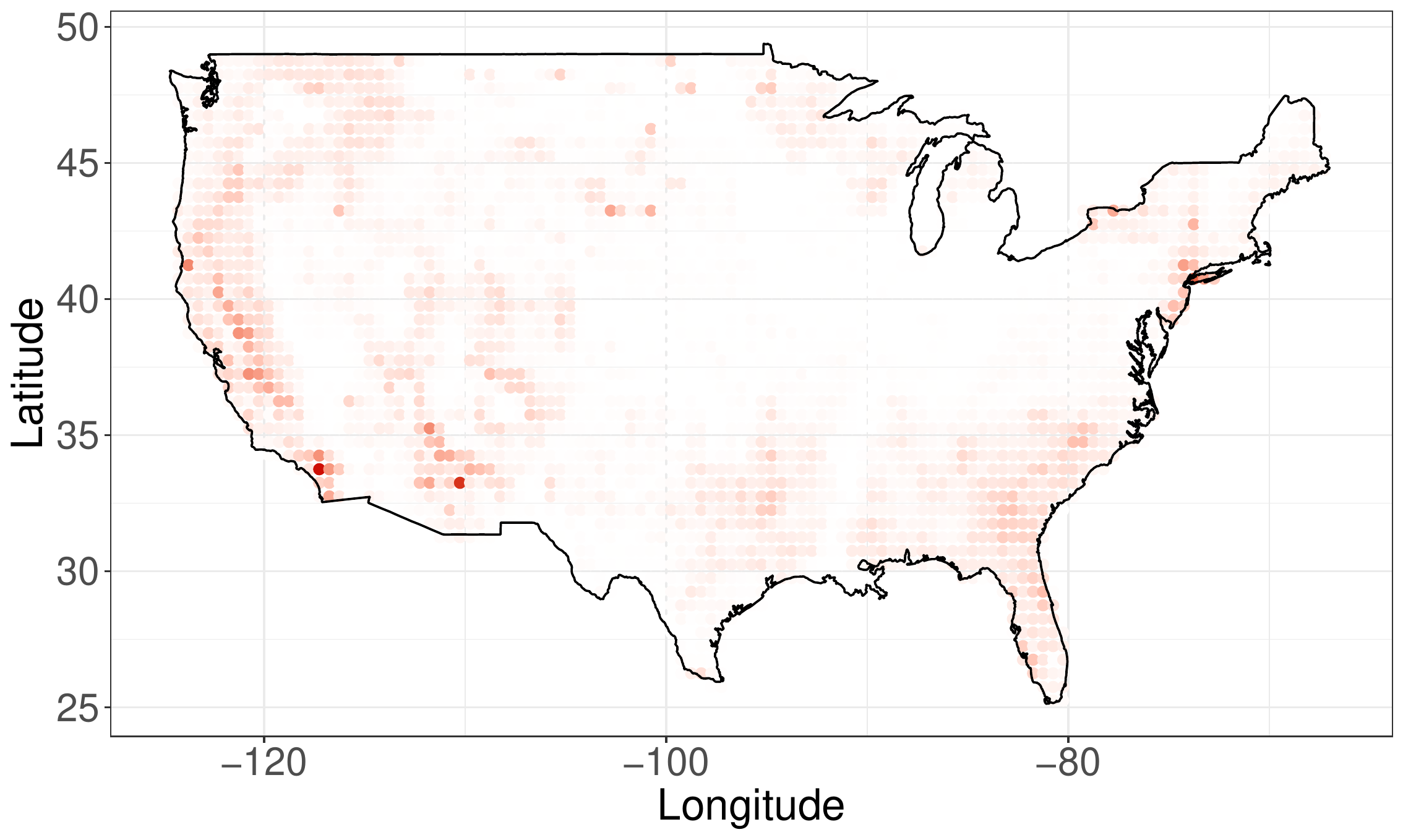}
    \caption{}
\end{subfigure}%
\begin{subfigure}{.5\textwidth}
    \centering
    \includegraphics[width=\textwidth]{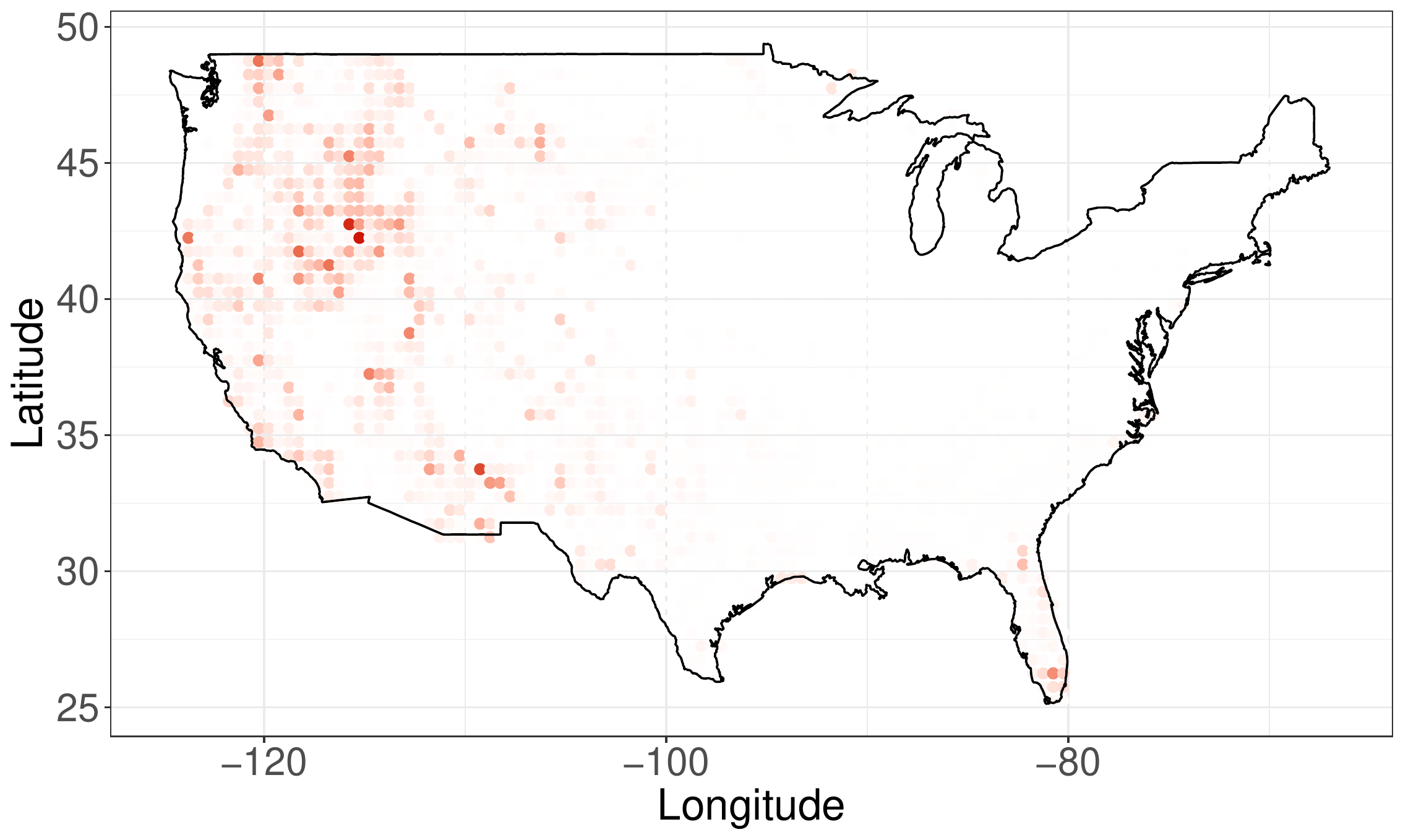}
    \caption{}
\end{subfigure}
\caption{Average CNT (a \& c) and BA (b \& d) across all years for each grid cell, for MAS (a \& b) and for MJJA (c \& d).}
    \label{fig:averageBACNT}
\end{figure}

Figure~\ref{fig:averageBACNT} shows the spatial distribution of average CNT and BA values in two different time groupings: in the summer months (May, June, July and August; MJJA), when wildfires are more likely to occur, and in the remaining cooler months (March, April and September; MAS). The highest average CNT values are observed in the east for MAS and the west for MJJA. The highest average BA values typically occur in the west of the US during MJJA whilst the majority of the eastern US locations have relatively low average BA values in both time groups, with the exception of Florida. This demonstrates that there is both spatial and temporal variability in the wildfire observations.

\begin{figure}[!htbp]
\centering
\begin{subfigure}{.5\textwidth}
    \centering
    \includegraphics[width=\textwidth]{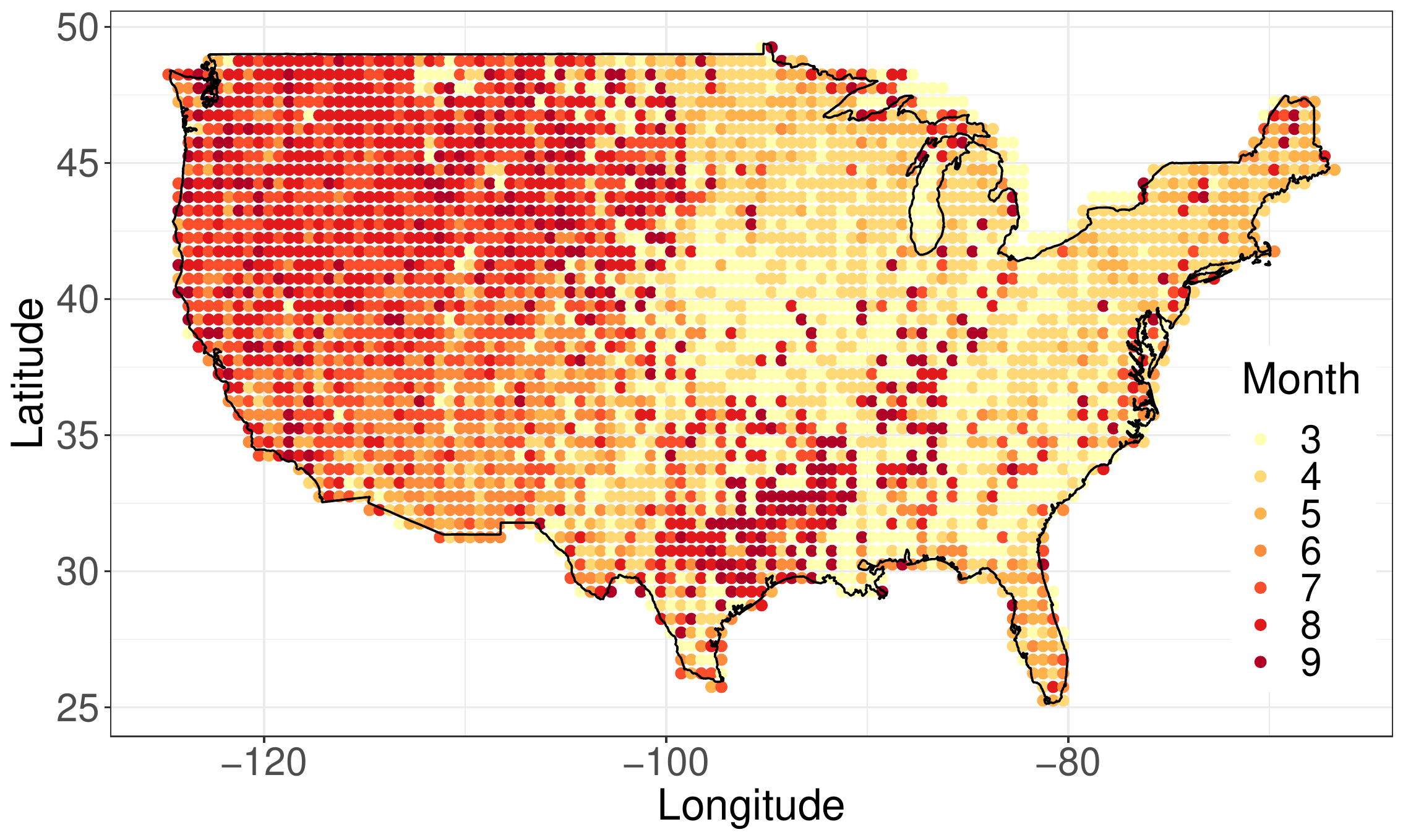}
    \caption{}
\end{subfigure}%
\begin{subfigure}{.5\textwidth}
    \centering
    \includegraphics[width=\textwidth]{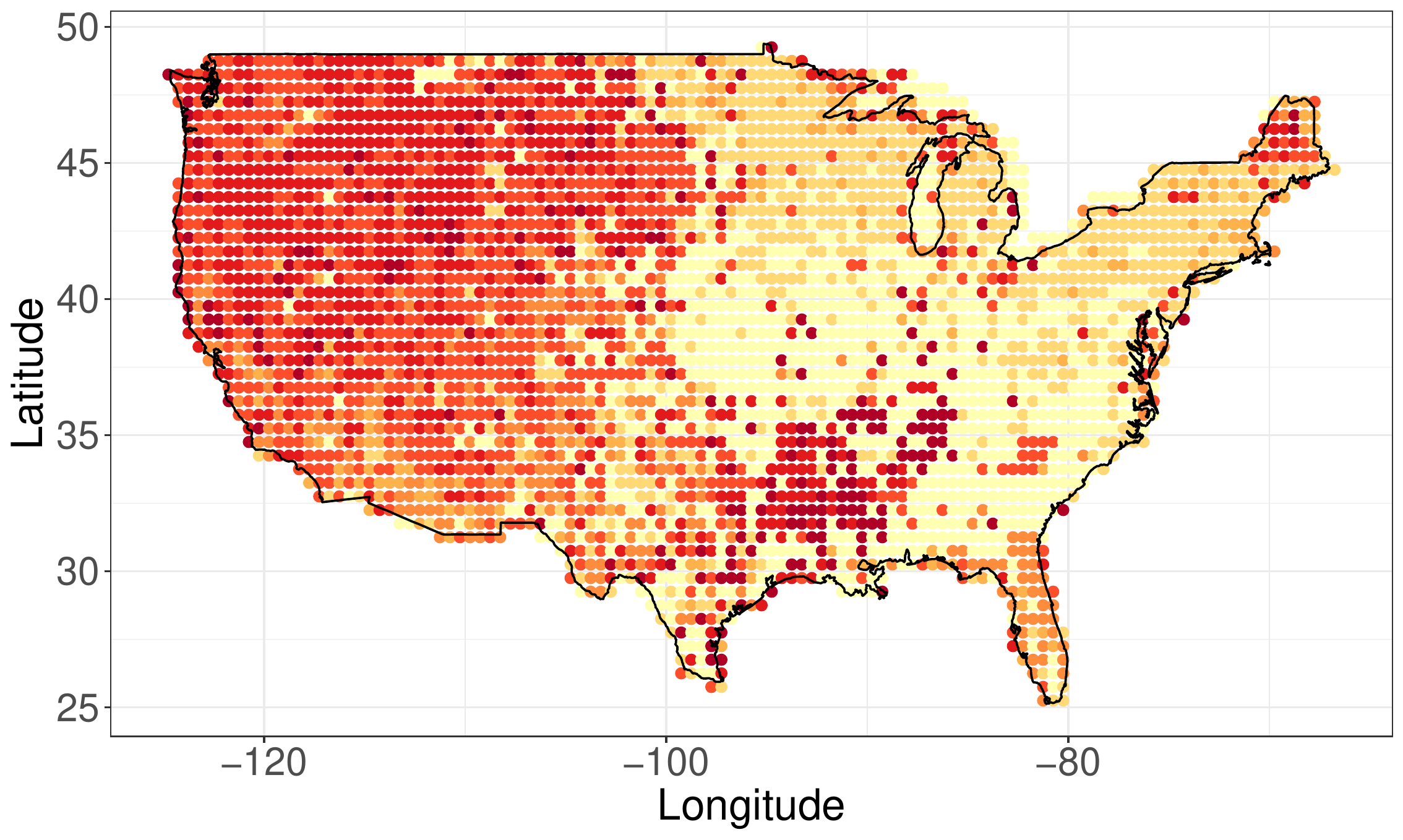}
    \caption{}
\end{subfigure}
    \caption{Month where the maximum CNT (a) and BA (b) across all years occurs for each grid cell.}
    \label{fig:maxmonBACNT}
\end{figure}

To further demonstrate this spatio-temporal variability, Figure~\ref{fig:maxmonBACNT} illustrates the months when the maximum CNT and BA observations occur for each grid cell. In the eastern US, the maxima of each variable tend to occur in July or August (shown by red points) whilst in the west, the maxima typically occur in March and April (illustrated by lighter yellow points). As global temperatures rise with anthropogenic climate change, the frequency and intensity of wildfires are generally expected to increase \citep{preisler2004,wuebbles2017climate}. To investigate this, we fit a linear model between year and annual mean CNT and BA separately, assuming independence across annual means. We find significant trends for both CNT and BA. Therefore, assuming stationarity across the entire spatial domain over the observation period would be unreasonable.

\begin{figure}[!ht]
\centering
\begin{subfigure}{.5\textwidth}
    \centering
    \includegraphics[width=\textwidth]{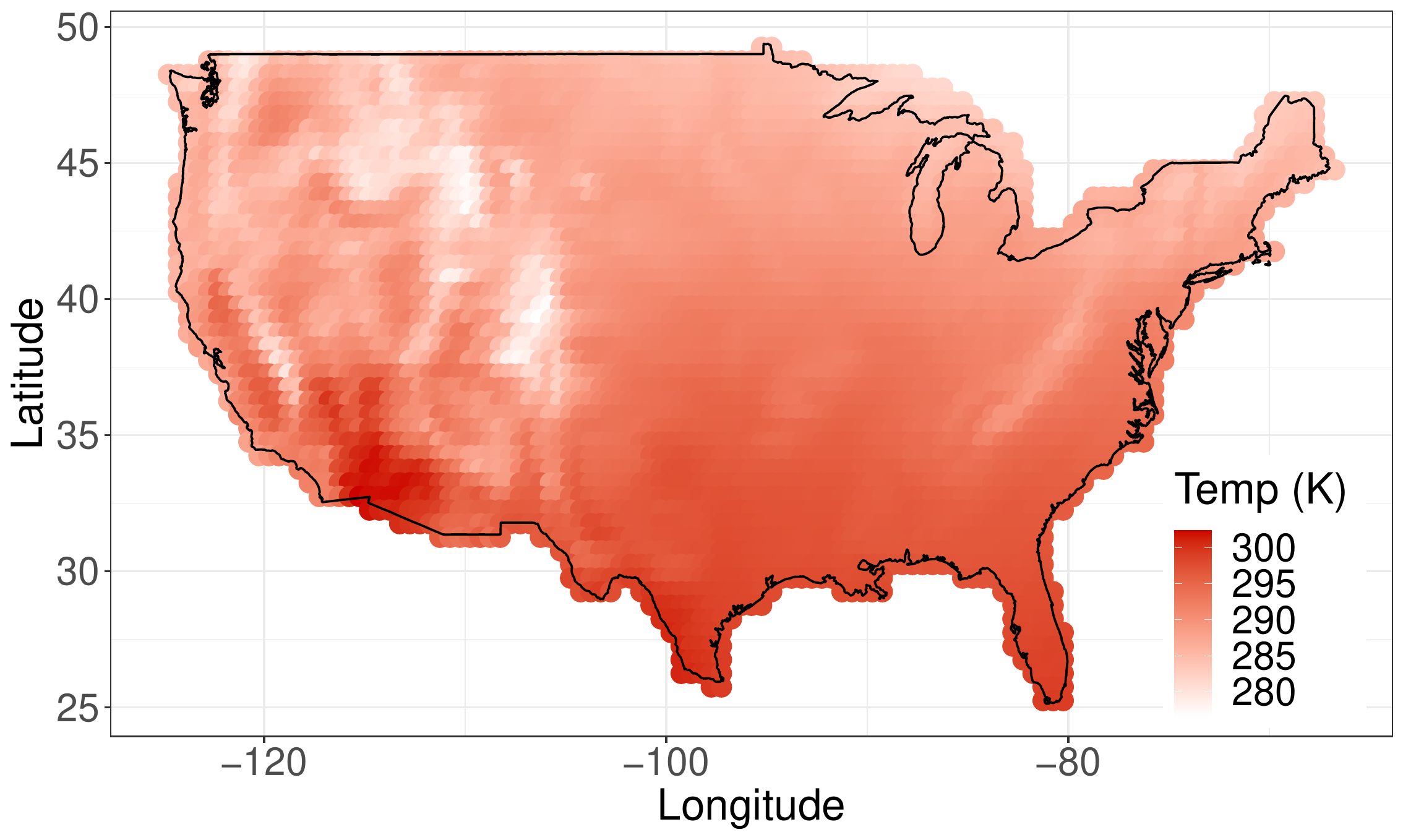}
    \caption{}
\end{subfigure}\begin{subfigure}{.5\textwidth}
    \centering
    \includegraphics[width=\textwidth]{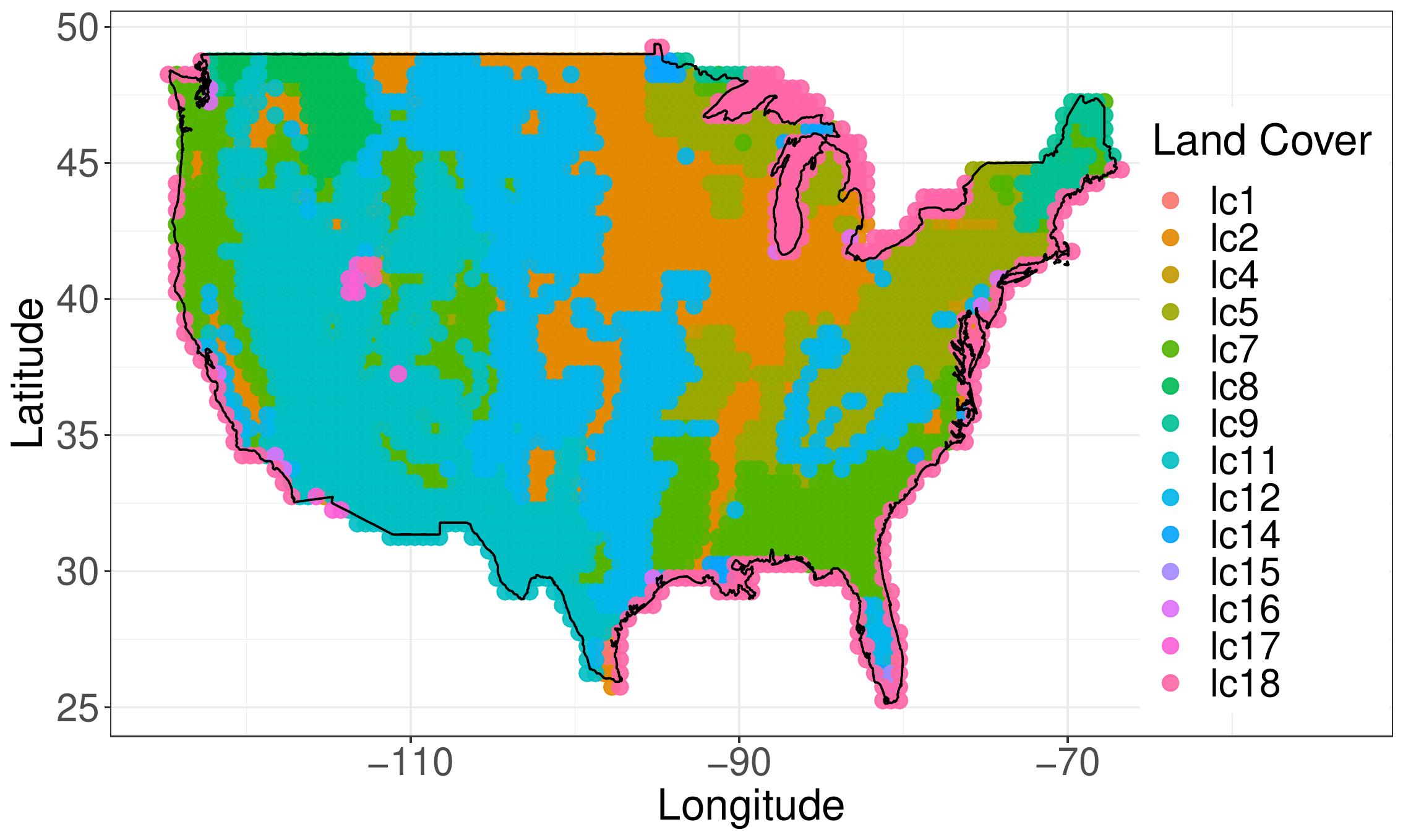}
    \caption{}
\end{subfigure}%
    \caption{Mean temperature in Kelvin (a) and the most common land cover variable (b) for each location.}
    \label{fig:max_lc}
\end{figure}

Due to the nature of wildfires, we expect to observe relationships between both CNT and BA observations and certain climate variables. For example, high temperature coupled with low rainfall and low wind speed are the ideal conditions for wildfires to ignite and spread \citep{holden2018decreasing,son2021changes}. No significant linear relationships exist for either wildfire variable with any of the climate covariates, suggesting such relationships are complex in nature. Figure~\ref{fig:max_lc}(a) shows the average temperature for each grid cell; temperature is non-stationary across the US but there is some spatial dependence, with nearby locations exhibiting similar values. Some form of spatial dependence exists for all climate variables. Since these variables are given as monthly averages, it is difficult to associate these covariates directly with the wildfire observations, which are also given as monthly aggregates.

Another factor likely to alter wildfire behaviour across the US is the type of land cover. For example, locations with large proportions of water or urban areas are typically not conducive to wildfires, whilst those with forest areas probably are. Eighteen land cover variables, given as proportions of each grid cell, are provided in the challenge data set; these are denoted $lc(j)$ for $j=1,\ldots,18$ and defined in~\cite{Opitz2021}. Figure~\ref{fig:max_lc}(b) illustrates the maximum land cover variable for each location. Spatial heterogeneity can be observed over different regions. For example, a large portion of the western US is taken up by shrubland ($lc(11)$), whereas the eastern region is dominated by cropland ($lc(j)$ for $j = 1, 2, 3$) and tree-based land cover types ($lc(j)$ for $j = 5, 6, 7, 8$). Unsurprisingly, many coastal locations are predominantly covered by water ($lc(18)$), and regions containing national forests (such as Kootenai and Stanislaus) are easily identifiable, since they are mostly made up of tree categories.

\subsection{Existing methods} \label{subsec:exist_method}
Various methods exist for modelling and predicting wildfire frequency and intensity. For example, generalised additive models (GAMs) with climatic, anthropogenic and/or spatial covariates are commonly used; see, e.g.,~\cite{Krawchuk2009}~or~\cite{Sa2018}. The latter captures covariate information via a fire index; many such indices have been proposed within the literature~\citep{ziel2020comparison}. Each index is typically developed with country-specific considerations in mind, such as land cover types and climate factors, and are often used by government bodies to assess risks and prioritise fire responses. In the US, the National Fire-Danger Rating System is the primary tool used for wildfire management~\citep{jack2002national}. There have been attempts in the literature to use fire indices as a means to model extreme wildfire events~\citep{koh2021spatiotemporal}. However, several approaches have found that certain fire indices are poor predictors of wildfires. For example, \cite{sharples2009simple} show that the Forest Fire Danger Index, typically used in Australia, is inadequate for predicting the behaviour of moderate to high-intensity wildfires. 

Machine learning techniques have also been adopted for wildfire modelling: \cite{Richards2022} and \cite{Ivek2022} use deep learning techniques; \cite{Cisneros2021} present a four-stage process including a random forest algorithm; and \cite{Koh2021} develops a gradient boosting model trained with loss functions appropriate for predicting extreme values. We take a simpler, marginal-based approach. 

The remainder of this paper is structured as follows. In Section \ref{sec:exploit_prop}, we illustrate how certain properties of the training data can be exploited to infer a subset of probability estimates for observations in the validation set. In Section~\ref{sec:marg_modelling}, we introduce our marginal modelling techniques for both CNT and BA. We also discuss our technique for estimating spatial neighbourhoods and corresponding tuning parameters. We conclude with a discussion of our approach in Section~\ref{sec:discussion}.

\section{Exploiting properties of the training data set} \label{sec:exploit_prop}

\subsection{Re-scaling burnt area values} \label{subsec:limit_BA} 
In this section, we discuss various properties of the wildfire data set, and how these can be exploited to improve the estimation of the predictive distributions for missing observations.  

To begin, observe that BA is an absolute measurement; this results in varying measurement scales across different locations. To better understand this, consider that some grid cells in the data set do not lie completely inside the continental US; this feature is captured by the `area' variable, denoted $p_i$, $i = 1, \hdots, N$, which describes the proportion of each grid cell that lies in the region of interest. BA observations depend upon this variable since for grid cells with smaller area values, there is less available land for wildfires to occur and hence lower BA values. For these reasons, the raw BA observations cannot be easily compared across locations. 

To account for this, we propose re-scaling BA observations to ensure all observations are on a unified, relative scale. Recall that $BA_i$, $i=1,\dots,N$, with $N=563,983$, denotes the $i$-th observation of the BA data, and that $BA^{val}\subset\{1,\dots,N\}$ is the set of indices for missing BA observations. We consider here the $i$-th observation, with corresponding grid cell area $p_i\in(0,1]$. For each $i\in\{1,\dots,N\}$, the total surface area of the grid cell is computed by taking the corresponding longitude and latitude coordinates and applying a formula derived from Archimedes' theorem \citep{kelly2021area}. We denote these surface area values by $SA_i$. The surface area contained within the continental US is then computed by multiplying the total surface area by the grid cell area variable, i.e., $SA_i\times p_i$. We denote these values by $SA^*_i$: such values will naturally vary between locations, especially for locations lying on a borderline. Moreover, $SA^*_i$ values naturally decrease going from South to North of the continent, since grid cells defined using longitude and latitude suffer from unequal cell sizes \citep{budic2016squares}. We refer to this variable as the true surface area. 

Using this variable, we derive a relative measure for BA, which we term burnt area proportion (BAP), i.e., for each $i$, define $BAP_i := BA_i/SA^*_i \in [0,1]$. This value denotes the proportion of the true surface area that has been burnt for each observation. It is arguably a better indicator of the impact and/or severity of wildfire events compared to raw BA observations, since it puts the absolute magnitude in context for each location. Moreover, this proportion is a relative measure, meaning the data for all locations are on the same scale; this allows for a more straightforward comparison between neighbouring observations with different (true) surface areas. 

We recall that the objective of the data challenge is to obtain probability estimates of the form $\Pr(BA_i \leq u)$ for all $u \in \mathcal{U}_{BA}$, where \begin{align*}
    \mathcal{U}_{BA}=\{&0, 1, 10, 20, 30, \hdots , 100, 150, 200, 250, 300, 400, 500, 1000, \\
& 1500, 2000, 5000, 10000, 20000, 30000, 40000, 50000, 100000\}, 
\end{align*} and $i \in BA^{val}$. This can be derived using the marginal distribution of $BAP_i$, since 
\begin{equation*}
    \Pr(BA_i \leq u) = \Pr(BAP_i\times SA^*_i \leq u) = \Pr(BAP_i \leq u/SA^*_i).
\end{equation*}
Consequently, we evaluate the distribution function of $BAP_i$ for all $u \in \mathcal{U}^i_{BAP}$, where $\mathcal{U}^i_{BAP}:= \mathcal{U}_{BA}/SA^*_i$, to obtain the required predictive probabilities. We introduce our technique for estimating this distribution function in Section~\ref{subsec:semiparametric_BA}.

We can also use these proportional data to deduce information about the upper tail of the distribution for $BAP_i$ at any $i \in BA^{val}$. Since it is impossible to observe a BA observation which exceeds the true surface area at any location, we can immediately deduce that $\Pr(BAP_i \leq u) = 1$ for any $u \in \mathcal{U}^i_{BAP}$ with $u\geq1$. In practice, over $1\%$ of missing BA observations satisfied the inequality $\max\{\mathcal{U}^i_{BAP}\}\geq1$, meaning a non-negligible amount of information can be uncovered via this preliminary step. 

We considered a similar re-scaling for CNT observations; however, there did not appear to be any obvious relationship between the true surface area and CNT values. Furthermore, unlike BA, no natural upper bound arises for CNT observations, so we cannot deduce properties of the upper tail distribution for missing observations.

\subsection{Exploiting features of the missing data}\label{subsec:dataFeatures}
Before introducing our marginal modelling procedures, we highlight how the training data can be used to provide information about the missing values we are required to estimate. This is possible since the missing values in the CNT and BA variables do not always occur at the same space-time locations, although there is some overlap in their missingness. We show that if exactly one of the CNT or BA values is known at a particular index, we can deduce information about the other. 

Recall that we are interested in estimating the predictive distribution of $CNT_i$ for some $i\in CNT^{val}$, i.e., $\Pr(CNT_i\leq u)$ for $u\in\mathcal{U}_{CNT}$, where $u\in\mathcal{U}_{CNT}=\{0,1,\dots,9,10,12,\dots,30,40,\dots,100\}$. If $i\not\in BA^{val}$ and $BAP_i=0$, we can immediately deduce that $CNT_i=0$ and $\Pr(CNT_i\leq u) = 1$ for all $u\in \mathcal{U}_{CNT}$. Moreover, if $i\not\in BA^{val}$ and $BAP_i>0$, we have that $CNT_i>0$, implying $ \Pr(CNT_i\leq 0) = 0$, though we are still required to estimate the predictive distribution for all $u\in\mathcal{U}_{CNT}\backslash \{0\}$. The values we can infer for $BAP_i$ from $CNT_i$, with $i\in BA^{val}$, are analogous, so the detail is omitted here.

We find that $CNT_i=0$ for approximately 23\% of the points in the CNT validation set, and that $CNT_i>0$ for an additional 15\%. We can also deduce similar proportions for the BAP values we are required to predict. A reasonable amount of information can therefore be uncovered using this simple step.

We also found that for the non-missing CNT and BA observations, the probability of observing a zero observation exceeded $0.999$ for both variables when $lc(18)_i>0.94$, where $lc(18)_i$ denotes the proportion of each location covered by water. Therefore, for any $i \in CNT_{val}$ ($i \in BA_{val}$) with $lc(18)_i>0.94$, we set $CNT_i = 0$ ($BAP_i = 0$), implying $\Pr(CNT_i\leq u) = 1, \forall~u\in \mathcal{U}_{CNT}$ ($\Pr(BAP_i\leq u) = 1, \forall~u\in \mathcal{U}^i_{BAP}$). 

As well as improving estimates of the predictive distribution of some locations, the additional steps introduced in this section also increase the amount of information available. This aids the marginal estimation procedures detailed in Sections~\ref{subsec:parametric_CNT} and~\ref{subsec:semiparametric_BA}. 

\section{Marginal modelling of missing values} \label{sec:marg_modelling} 

\subsection{Neighbourhood selection} \label{subsec:neighbourhood_select}

For our approach, we make the following assumption: for any observation with index $i\in\{1,\dots,N\}$, there exists some spatial neighbourhood of indices, $\mathcal{N}_i$, where all corresponding observations come from the same marginal distribution. Through estimation of this distribution, we can obtain predictive probabilities for missing CNT and BAP observations. In this section, we introduce our approach for selecting these neighbourhoods for CNT observations; the approach for BAP is analogous. 

Consider the observation with index $i\in\{1,\dots,N\}$, and denote the corresponding spatial location, month and year by $\bm{s}_i\in\mathbb{R}^2$, $m_i\in\{3,\ldots,9\}$ and $y_i\in\{1993,\ldots,2015\}$, respectively. We define the spatial neighbourhood as
\begin{align}
    \mathcal{N}_i := \left\{j\in\{1,\dots,N\} : \|\bm{s}_i-\bm{s}_j\|\leq k_1^{CNT} , m_j=m_i, y_j=y_i\right\},
\label{eqn:spatialNeighbourhoods}
\end{align}
for some $k_1^{CNT}\geq 0$, i.e., the indices of all observations occurring in the same year and month as observation $i$ with a spatial distance of at most $k_1^{CNT}$ from $\bm{s}_i$. The spatial distance $\|\cdot\|$ is measured in kilometres (km) using the Haversine formula; in practice, these are calculated via the \texttt{distm} function in the \texttt{R} package \texttt{geosphere} \citep{geosphere}. We treat $k_1^{CNT}$ as a tuning parameter and introduce a cross validation technique to select it in Section~\ref{subsec:validation}. We denote the $CNT$ values corresponding to neighbourhood $\mathcal{N}_i$ by $CNT^{\mathcal{N}_i}= \left\{CNT_j:j\in\mathcal{N}_i\right\}$. The definitions of $k_1^{BAP}$ and $BAP^{\mathcal{N}_i}$ for $i\in\{1,\dots,N\}$ are analogous. 

More complex spatial neighbourhoods, which incorporated temporal and covariate-based information, were also considered but ultimately resulted in worse quality marginal estimates. This is discussed in the Appendix, where we present prediction scores for other neighbourhoods we considered. 

\subsection{A parametric approach for modelling CNT}\label{subsec:parametric_CNT} 
Following \cite{joseph2019}, we assume all observations in the set $CNT^{\mathcal{N}_i}$ follow a zero-inflated negative binomial distribution for all $i\in CNT^{val}$, i.e., for any $CNT \in CNT^{\mathcal{N}_i}$, we have that 
\begin{equation} \label{eqn:ZINB}
    \Pr (CNT = j) = \begin{cases}
        \pi + (1-\pi)g(0) \hspace{1em} &\text{if} \; j=0, \\
        (1-\pi)g(j) \hspace{1em}&\text{if} \; j > 0, 
    \end{cases}
\end{equation}
where $\pi \in [0,1]$ denotes a probability and $g(j), \;j \geq 0$, is the probability mass function of the negative binomial distribution. We estimate the parameter $\pi$ and those of the negative binomial distribution using likelihood inference. We then evaluate distribution~\eqref{eqn:ZINB} for all $u\in\mathcal{U}_{CNT}$ using the estimated parameters, resulting in the predictive distribution for the missing observation $CNT_i$. In practice, we use the same tuning parameter, $k_1^{CNT}$, for all $i \in CNT^{val}$; we discuss our approach to selecting this value in the Section~\ref{subsec:validation}. 

\subsection{A semi-parametric approach for modelling BAP}\label{subsec:semiparametric_BA}

Given any $i\in BA^{val}$, we assume all observations in the set $BAP^{\mathcal{N}_i}$ follow the semi-parametric marginal distribution given in \cite{Richards2021}. This distribution was proposed for modelling precipitation data, which are similar to wildfire data in the sense that they typically contain a large number of zero observations. These data structures are referred to as mixture distributions, since they are a mix of a discrete (zero observations) and a continuous (positive BAP observations) process. Values in the bulk of the data, including zeros, are modelled empirically, while values in the upper tail are modelled using a generalised Pareto distribution (GPD). This distribution is typically referred to in the context of the `peaks over threshold' approach \citep{Balkema1974,pickands1975statistical}, whereby a GPD is fitted to independent and identically distributed exceedances of a high threshold. This overall marginal model is given by
\begin{equation} \label{eqn:SPD}
        \Pr(BAP \leq x) = \begin{cases} 
                   z_i \hspace{1em} & \text{if} \; x = 0,\\
                   \frac{1 - \lambda_i - z_i}{F^*_{i}(u_i)} F^*_{i}(x) + z_i \hspace{1em} & \text{if} \; 0 < x \leq u_i, \\
                   1 - \lambda_i (1-H_{u_i}(x)) \hspace{1em} & \text{if} \; x > u_i,
                \end{cases}
\end{equation}
for all $BAP \in BAP^{\mathcal{N}_i}$, where $z_i$ is the probability of observing a zero, $u_i$ is some high threshold to be chosen, $\lambda_i = \Pr(BAP>u_i)$, $F^*_{i}$ is the distribution function of strictly positive observations, and $H_{u_i}(x)$ denotes the cumulative distribution function of the GPD, i.e., $H_{u_i}(x) = 1- \left[ 1 + (\xi_i(x-u_i))/(\sigma_i) \right]_+^{-1/\xi_i}$, with $x_+=\max\{x,0\}$ and $(\sigma_i,\xi_i)\in\mathbb{R}_+\times\mathbb{R}$. We refer to $\sigma_i$ and $\xi_i$ as the scale and shape parameters, respectively. See \cite{Coles2001} for a more detailed discussion of the peaks over threshold approach. 

We set $k_2^{BAP} := 1 - \lambda_i$ for all $i \in BA^{val}$ and treat $k_2^{BAP}$ as another tuning parameter, which we again estimate using cross validation; see Section~\ref{subsec:validation}. Both $u_i$ and $z_i$ can be estimated empirically, alongside $F^*_{i}$. Note that this marginal model is only valid when $z_i < 1-\lambda_i$: in such cases, the GPD scale and shape parameters are estimated using likelihood inference. We then evaluate the distribution described in equation~\eqref{eqn:SPD} at the fitted parameters for all $u\in\mathcal{U}^i_{BAP}$, resulting in the predictive distribution for the missing observation $BAP_i$.

In the cases when $z_i \geq 1-\lambda_i$ (i.e., the marginal model is not valid), we use a fully empirical distribution. Such cases occur when the estimated threshold equals zero, corresponding to neighbourhood sets containing a significant proportion of zeros, indicating a low occurrence of wildfires. 

\subsection{Tuning parameter selection}\label{subsec:validation}
\begin{figure}
    \centering
    \includegraphics[width=0.5\textwidth]
    {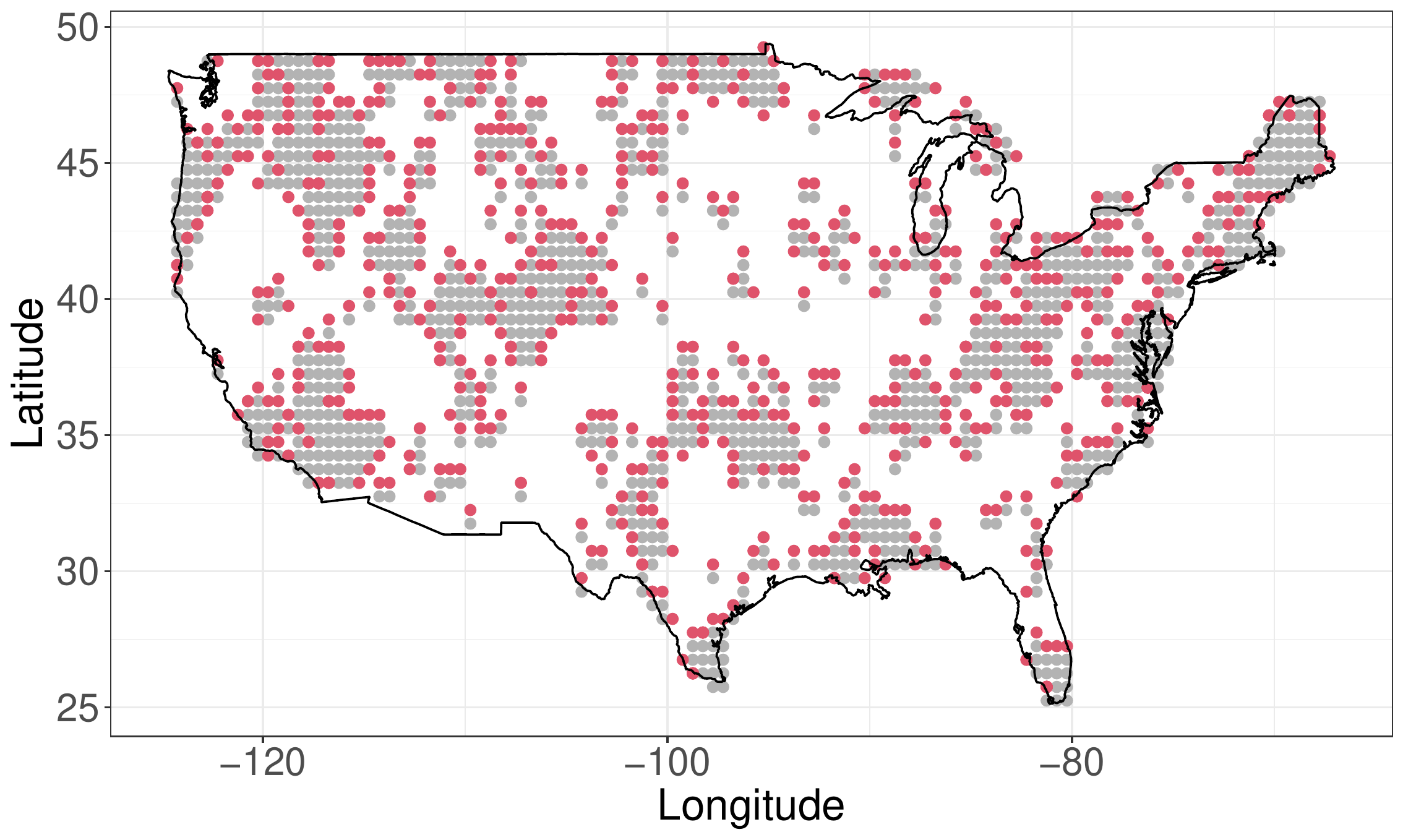}
    \caption{Locations in the set $CNT^{val}$ for March 1994 (grey) and the corresponding locations of observations for tuning parameter selection (red).}
    \label{fig:validationLocations}
\end{figure}

We now consider how to select the tuning parameters $k_1^{CNT}, k_1^{BAP}$ and $k_2^{BAP}$ used in our marginal modelling approaches. One option is to use leave-one-out cross validation and select the tuning parameter values that minimise the score used for ranking in the data challenge: see \cite{Opitz2021} for more information. However, the locations for the validation data are not randomly distributed across the spatial domain, and are generally clustered in space and time. We demonstrate this in Figure~\ref{fig:validationLocations}, where we show the locations of the CNT validation data for March 1994; the resulting plots have similar features for BAP, as well as for different months and years. In the case of BAP data, this implies that for a fixed value of $k_1^{BAP}$, there will be a larger number of missing values in the set $BAP^{\mathcal{N}_i}$ for $i\in BA^{val}$ than for $i\not\in BA^{val}$, on average. The same holds when considering CNT data for a fixed value of $k_1^{CNT}$. This feature of the validation set means that using standard leave-one-out cross validation over all training locations could lead to selecting smaller neighbourhoods than are really appropriate. 

We instead propose to carry out the parameter selection procedure using only a subset of the observations in the training data. Focusing on BAP, for each $i\in BA^{val}$ and any combination of $(k_1^{BAP},k_2^{BAP})$ values, we allow the observation indexed by
\[
\argmin\limits_{j:m_i=m_j, y_i=y_j, j\not\in BA^{val}} \|\bm{s}_i-\bm{s}_j\|,
\]
to contribute to the score, i.e., giving the spatially-nearest non-missing observation that occurs in the same month and year as observation $i$. Ties may be broken at random, or using any rule that results in only one nearest neighbour per location. Since these locations can be the nearest neighbour of more than one validation location, some of the corresponding observations are included more than once in the score calculation. An alternative would have been to use each of these observations only once to avoid duplicates, but this means some validation locations would not be represented in the score calculation. 

We consider the following candidate values for the tuning parameters: $k_1^{BAP}\in\{50,75,\dots,400\}$, $k_2^{BAP}\in\{0.05,0.10,\dots,0.95\}$. For each combination of candidate values, we recalculate the score function proposed in \cite{Opitz2021}, summing over all values corresponding to our set of nearest neighbours, before finally selecting the parameter combination that minimises the score. The procedure in the CNT case is analogous, albeit without the GPD quantile parameter $k_2^{BAP}$. In Figure~\ref{fig:validationLocations}, we demonstrate the locations of observations that contribute to the tuning parameter selection procedure for CNT in March 1994. This results in selected tuning parameter values of $k_1^{BAP}=175$, $k_2^{BAP}=0.5$, and $k_1^{CNT}=125$. 

We note that our final approach has similarities with the winning entry to the 2017 EVA data challenge \citep{Stephenson2018}, where the authors combine data across locations with sufficient observations, in order to fit a generalised extreme value distribution for predicting precipitation extremes. They advocate the use of cross validation for tuning parameter selection and to compare potential modelling approaches in data challenges such as this, where the aim is to optimise some pre-determined metric.

\section{Discussion of limitations and possible extensions}\label{sec:discussion}
In this paper, we have discussed a marginal modelling approach for predicting wildfire events across the contiguous US. This framework was applied to obtain estimates of the cumulative distribution function at locations with missing entries for either CNT or BA. The resulting estimates were then ``ranked'' using a score function weighted to give higher importance to extreme observations \citep{Opitz2021}. Our method produced scores of $4080.559$ and $3640.92$ for CNT and BA, respectively, resulting in an overall score of 7721.479; this is a significant improvement on the proposed benchmark technique.  

Unlike all the techniques introduced in Section \ref{subsec:exist_method}, our approach does not attempt to specify the relationships between the auxiliary and wildfire variables. Such relationships appear to be complex and non-linear in nature, which may be explained by a variety of hypotheses. For example, the monthly aggregated format of the wildfire variables arguably makes it more difficult to associate them with any climate covariates, which are given as monthly means. Instead, our approach relies on the assumption that wildfire observations within spatial neighbourhoods arise from the same marginal distribution. We believe that this is realistic since neighbouring locations are likely to have similar auxiliary covariates, as demonstrated in Figure \ref{fig:max_lc}, and a large wildfire event occurring at one location is likely to increase the probability of wildfires in neighbouring locations. Furthermore, since the values in neighbourhood sets vary over time for each missing CNT or BA observation, our approach accounts for the temporal non-stationarity discussed in Section \ref{subsec:data_exploration}. We also propose several preliminary steps in Section \ref{sec:exploit_prop}; these steps do not require expert knowledge of wildfires to implement. Furthermore, such steps lead to significant improvements in the predictive distributions obtained using our approach by increasing the amount of information available and bringing all BA observations onto a unified scale. 

While developing our approach, we investigated the possibility of accounting for covariate influence on extreme CNT values via the use of GAMs (this option was also mentioned in Section~\ref{subsec:exist_method}), but found their predictive performance to be poor in this setting; see \cite{Wood17} for details on these types of models. In particular, we fitted a continuous GPD for each month, with the scale parameter having a GAM form \citep{Yng19} comprising spatial and climatological covariates; a continuous approximation was used due to having discrete CNT values. The advantage of this method is that covariate effects can be directly assessed by examining the smooth functions underlying the models. From the fitted GAMs, the general spatial behaviour of the CNT data was modelled fairly well in each month, but these models did appear to suffer from oversmoothing, even when using models with the lowest level of smoothness. On the other hand, we found that physically-interpretable covariate behaviour for the climatological variables was hard to capture, making model selection difficult; the precise reason for this is unclear. It is likely that the aforementioned oversmoothing combined with other issues, such as poor convergence of the underlying numerical optimisation routines and difficulties combining the fitted GAMs with models for the bulk of the CNT values, lead to poor model performance against the benchmark. Therefore, it appears that this type of approach may not be favourable in situations where the prediction of unknown values is required, and is more suited to analyses where the aim is to account for uncertainty whilst modelling complex covariate effects. Indeed, in addition to the GAM-based approach mentioned in Section~\ref{subsec:exist_method}, \cite{Sa2018}, \cite{zhang2017wildfire} and \cite{rodriguez2020evaluating} have also successfully applied GAM techniques for modelling wildfire data.

One possible extension of the modelling techniques proposed in Sections \ref{subsec:parametric_CNT} and \ref{subsec:semiparametric_BA} would be to introduce weights into the marginal estimation procedures. In the current format, observations within spatial neighbourhoods are given equal weights, even though it is likely that locations with a closer proximity to a missing observation would provide more useful information than locations that are further away. Our current method could therefore be extended by introducing weights to the marginal estimation procedures, with closer observations contributing more to probabilistic estimates. We would expect different values of the tuning parameters $k_1^{CNT}$ and $k_1^{BAP}$, defining the spatial range of the neighbourhoods, to be appropriate in this case, but our cross validation approach could be used analogously. 

Changes to the definition of the neighbourhoods in equation~\eqref{eqn:spatialNeighbourhoods} could lead to improvements with our approach. One drawback with our current implementation is that the values $k_1^{CNT}$ and $k_1^{BAP}$ are chosen to be the same across all validation locations. Although we selected these tuning parameters carefully via cross validation, it is possible that this is an over-simplification and allowing the values to depend on covariates such as location, month or year may have been more appropriate. An extension of our approach could allow for this possibility, e.g., by separating the spatial domain into smaller sections and implementing cross validation separately in each one. It may also be reasonable to apply clustering algorithms as a preliminary step, to inform the spatial regions where setting the tuning parameters $(k_1^{CNT},k_1^{BAP})$ as constant is a reasonable assumption. Allowing these parameters to vary across space also has the potential to provide insight into the behaviour of wildfires across the spatial domain. Additionally, we considered allowing the neighbourhoods themselves to depend on covariate-based clusters or to cover larger time windows, but the results presented in the Appendix suggest the simpler spatial neighbourhood approach was more successful.

While the zero-inflated negative binomial distribution proposed for CNT neighbourhoods is not motivated by extreme value theory, our analysis indicated the fitted marginal distributions performed reasonably well, including in the upper tail in the majority of cases. Several other distributions were tested, including fully empirical and discrete GPD models \citep{HitzEA17}; however, in every case, these distributions resulted in poorer prediction quality when ranked by the objective function given in \cite{Opitz2021}. This is likely due to the difficulties that arise in trying to capture behaviour in the bulk and tail simultaneously, and perhaps due to an insufficient amount of data in each of our spatial neighbourhoods for fitting the discrete GPD. In addition, alternative marginal distributions for BAP observations have the potential to further improve the predictive ability of our modelling framework. 

\bmhead{Acknowledgments}
We are grateful to the referees and guest editor for constructive comments that have greatly improved the paper. 

\section*{Declarations}
\subsection*{Funding}
This paper is based on work completed while Callum Murphy-Barltrop and Eleanor D'Arcy were part of the EPSRC funded STOR-i centre for doctoral training (EP/L015692/1 and EP/S022252/1, respectively). Emma Simpson was partly supported by the King Abdullah University of Science and Technology (KAUST) Office of Sponsored Research (OSR) under Award No.\ OSR-2017-CRG6-3434.02.
\subsection*{Competing interests}
The authors have no relevant financial or non-financial interests to disclose.
\subsection*{Data availability}
The data set analysed during the current study is available from the corresponding author on reasonable request.

\begin{appendices}




\section{Spatio-temporal neighbourhoods}\label{sec:Appendix}
As alternatives to the spatial neighbourhoods $\mathcal{N}_i$, $i\in\{1,\dots,N\}$, defined in equation~\eqref{eqn:spatialNeighbourhoods}, we also considered neighbourhoods of the form
\begin{align*}
    \mathcal{N}^{t}_i := \big\{j\in&\{1,\dots,N\} : \|\bm{s}_i-\bm{s}_j\|\leq k_1^{CNT} , m_j=m_i, \\
    &y_j \in \{y_i - k_y^{CNT},y_i - k_y^{CNT} +1, \dots ,y_i + k_y^{CNT}\} \big\},
\end{align*}
and 
\[
\mathcal{N}^{c}_i := \left\{j\in\{1,\dots,N\} : \|\bm{s}_i-\bm{s}_j\|\leq k_1^{CNT} , m_j=m_i, y_j=y_i, c_j = c_i \right\},
\]
for some $k_1^{CNT} \geq 0$ and $k_y^{CNT} \in \mathbb{N}$, where $c_j$ denotes a covariate-based cluster assignment for each observation $j \in \mathcal{N}_i$. Analogous neighbourhoods were also considered for BAP. 

With the observation month fixed and the spatial range defined as in $\mathcal{N}_i$, the neighbourhood $\mathcal{N}^{t}_i$ incorporates additional observations from neighbouring years, thus increasing the amount of data available for marginal estimation and adding a temporal element to the modelling procedure. On the other hand, the month and year are fixed for $\mathcal{N}^{c}_i$ so that only those data points with the same cluster assignment as observation $i$ are considered, thus reducing the amount of information available for a fixed $k_1^{CNT}$ or $k_1^{BAP}$ value. However, assuming we can define clusters such that observations in the same cluster have more similar marginal tail properties, this additional step has the potential to improve marginal estimation.

Cluster assignments used within the $\mathcal{N}^{c}_i$ neighbourhoods were computed using divisive hierarchical clustering \citep{rokach2005clustering} for two different covariates: temperature and precipitation. We select these variables since they have been shown to be positively and negatively associated, respectively, to wildfire events \citep{duane2021towards,crockett2018greater}, and therefore may allow us to group together locations with similar marginal properties for CNT and BAP. 

We use hierarchical clustering since this technique has been used in practice to approximate spatial clusters with similar wildfire properties \citep{rodrigues2019spatial,rodrigues2019identifying,rahimi2020comparative}. The clustering procedure is as follows: 
\begin{enumerate}
    \item Standardise the auxiliary variable  data (temperature or precipitation) for every location in $\mathcal{N}_i$. 
    \item Apply hierarchical clustering, using the standardised covariate data, to obtain two clusters, i.e., for each $j \in \mathcal{N}_i$, $c_j=1$ or $2$.
    \item Compute the subset of locations with the same cluster assignment as location $i$, i.e., $\{j \in \mathcal{N}_i \mid c_j = c_i\}$. 
\end{enumerate}

In our analysis, we found that both the neighbourhoods $\mathcal{N}^{t}_i$ and $\mathcal{N}^{c}_i$ resulted in worse prediction scores compared to the simpler spatial neighbourhood approach outlined in Section \ref{sec:marg_modelling}. This is illustrated by the results in Tables \ref{table:time_scores} and \ref{table:cluster_scores}, where we present the overall prediction scores, as outlined in \cite{Opitz2021}, for $\mathcal{N}^{t}_i$ and $\mathcal{N}^{c}_i$, respectively; recall that we aim to minimise this score. 

For these scores, we let $k_1^{CNT}= k_1^{BAP}:=k_1\in \{50,75,\dots,250\}$ to incorporate a variety of spatial distances and set $k_2^{BAP}=0.5$ to match the existing selected tuning parameter from Section~\ref{subsec:validation}. For $\mathcal{N}^{t}_i$, we let $k_y^{CNT}=k_y^{BAP}:=k_y \in \{1,2,3,4,5,6\}$, resulting in time windows of up to 13 years. Note that these scores correspond to the final prediction scores, i.e., when the missing data are known, and in practice, one would need to select the tuning parameters using the cross validation procedure outlined in Section~\ref{subsec:validation}.

\begin{table}[!h]

\caption{Total scores obtained using neighbourhood $\mathcal{N}^{t}_i$.\label{table:time_scores}}
\centering
\begin{tabular}[t]{ccccccc}
\hline
Distance & $k_{y}$ = 1 & $k_{y}$ = 2 & $k_{y}$ = 3 & $k_{y}$ = 4 & $k_{y}$ = 5 & $k_{y}$ = 6\\
\hline
$k_{1}$ = 50 & 8393.3 & 8089.3 & 8003.6 & 7991.2 & 7925.9 & 7909.3\\

$k_{1}$ = 75 & 8225.5 & 8134.2 & 8161.9 & 8187.4 & 8162.6 & 8159.6\\

$k_{1}$ = 100 & 8295.6 & 8222 & 8264.4 & 8293.9 & 8275.5 & 8274.1\\

$k_{1}$ = 125 & 8439.9 & 8401.4 & 8457 & 8487.2 & 8474.1 & 8474\\

$k_{1}$ = 150 & 8547 & 8519 & 8578.2 & 8608.6 & 8599.5 & 8600.6\\

$k_{1}$ = 175 & 8624.8 & 8604.8 & 8664.5 & 8694.1 & 8687.5 & 8689\\

$k_{1}$ = 200 & 8711.2 & 8696.7 & 8755.2 & 8784.2 & 8778.4 & 8781.5\\

$k_{1}$ = 225 & 8779.8 & 8768.9 & 8823.8 & 8852.3 & 8848 & 8851.2\\

$k_{1}$ = 250 & 8843.9 & 8838.3 & 8892 & 8919.1 & 8915.4 & 8919.4\\
\hline
\end{tabular}
\end{table}

\begin{table}[!h]

\caption{Total scores obtained using neighbourhood $\mathcal{N}^{c}_i$ with clusters computed using temperature and precipitation.
\label{table:cluster_scores}}
\centering
\begin{tabular}[t]{ccc}
\hline
Distance & Temperature clusters & Precipitation clusters\\
\hline
$k_{1}$ = 50 & 11149 & 11161\\

$k_{1}$ = 75 & 9608.8 & 9607.7\\

$k_{1}$ = 100 & 9245.2 & 9309.2\\

$k_{1}$ = 125 & 8651.9 & 8718.5\\

$k_{1}$ = 150 & 8522.9 & 8582.1\\

$k_{1}$ = 175 & 8431.5 & 8505.2\\

$k_{1}$ = 200 & 8460.3 & 8489.5\\

$k_{1}$ = 225 & 8470.2 & 8511.1\\

$k_{1}$ = 250 & 8493.6 & 8539.5\\
\hline
\end{tabular}
\end{table}

One can observe that all the prediction scores from Tables \ref{table:time_scores} and \ref{table:cluster_scores} exceed the final score obtained using the method described in Section \ref{sec:marg_modelling}, and in many cases, the scores obtained were significantly worse. The fact that these predictions were worse for a wide range of tuning parameter combinations gives further support to our main modelling approach. 

In the case of temporal neighbourhoods, since the predictive scores do not tend to decrease with the parameter $k_y$, our results suggest that marginal wildfire behaviour can vary significantly over neighbouring years.

Therefore, even though incorporating information from neighbouring years increases the amount of data available for model fitting, it does not appear to improve the quality of marginal estimates. 

In the case of the cluster-based neighbourhoods, our results indicate that observations with similar temperature and precipitation values may not be those with similar wildfire behaviour. We suspect this may occur due to the complex nature of the relationships between the wildfire and auxiliary variables described in Section \ref{sec:discussion}. Such relationships are unlikely to be picked up by incorporating this additional clustering step. Clustering also reduces the amount of data available for model fitting, which also appears to reduce the quality of marginal estimates. 

On the whole, these results suggest that incorporating additional information, both from temporal windows and covariate-based clusters, does not improve the quality of marginal estimates for either CNT or BAP under our modelling approach. Combined with the principle of parsimony, we do not consider these alternative neighbourhoods further. 

\end{appendices}


\bibliography{sn-bibliography}


\end{document}